\documentstyle[12pt,aasms]{article}
\tightenlines

\begin{document}

\title {Aromatic Hydrocarbons, Diamonds, and Fullerenes
in Interstellar Space: 
Puzzles to be Solved by Laboratory and Theoretical
Astrochemistry}

\bigskip 

\author {K. Sellgren$^1$}

\bigskip 

\affil
{Astronomy Department, Ohio State University,
140 West 18th Avenue, Columbus, OH  43210  USA }

\altaffiltext{1}
{Corresponding author: Tel.: 614--292--1898;
fax: 614--292--2928; e-mail address: sellgren@astronomy.ohio--state.edu}

\bigskip

\centerline {\bf Abstract}

New research is presented, and previous research is reviewed,
on the emission and absorption of interstellar aromatic hydrocarbons.
Emission from aromatic hydrocarbons dominate the 
mid-infrared emission of many galaxies,
including our own Milky Way galaxy.
Only recently have aromatic hydrocarbons been observed in absorption
in the interstellar medium, along lines of sight with high column densities
of interstellar gas and dust.
Much work on interstellar aromatics has been done, with astronomical
observations and laboratory and theoretical astrochemistry.
In many cases the predictions of laboratory and theoretical work are
confirmed by astronomical observations, but in other cases clear discrepancies
exist which provide problems to be solved by a combination of astronomical
observations, laboratory studies, and theoretical studies.
The emphasis of this paper will be on current outstanding puzzles 
concerning aromatic hydrocarbons which
require further laboratory and theoretical astrochemistry to resolve.
This paper will also touch on related topics where
laboratory and theoretical astrochemistry studies are needed to
explain astrophysical observations, such as a
possible absorption feature due to interstellar ``diamonds'' and the 
search for fullerenes in space.

\section {Introduction}

Interstellar atomic gas, molecular gas, and dust grains,
where grains are solid particles with sizes of 1 -- 1000 nm,
pervade the space in between the stars known as
the interstellar medium (ISM).
The grains and molecules
in the lower density regions of our Galaxy, called
the diffuse ISM, are believed to form 
primarily in the outflowing gas from 
dying stars with masses similar to that of our own Sun. 
The final stage of this type of stellar death is a shell of
ejected gas, known as a planetary
nebula, which surrounds the stellar corpse, known as a white dwarf.
Grains and molecules find it difficult to form in the general
ISM, which has typical gas densities of 1 -- 10 
atoms cm$^{-3}$, and thus extremely low chemical reaction rates.
The outflows from dying stars have much higher densities
($>$ 10$^6$ atoms cm$^{-3}$), which make chemical reactions leading to
formation of molecules and grains much more probable.

In our universe, the pattern of cosmic elemental abundances is determined
first by the hydrogen and helium formed in the Big Bang, 
when our universe began,
and secondly by nuclear reactions within stars, which account for 
virtually all the other elements.
As a result of the pattern of common nuclear reactions within stars, 
only a very restricted set of elements are important in
interstellar chemistry.
Figure 1 shows the cosmic abundance, relative to hydrogen, 
of all elements which
have an abundance larger than 10$^{-6}$ relative to hydrogen [1].
It can be seen that the most important elements for interstellar
chemistry are H, O, C, N, Mg, Si, Fe, S, Al, Ca, Na, and Ni.
The inert gases He, Ne, and Ar are also relatively abundant but do
not interact chemically.
Other elements exist in the universe, but their scarcity 
leads to their playing a minor role in interstellar chemistry.
By far the most abundant interactive elements are 
H, O, C, and N, and thus it is
no surprise that the overwhelming majority of interstellar molecules
detected to date are primarily composed of these elements.

The interstellar infrared emission features (IEFs), at 3.3, 6.2, 7.7, 8.6,
11.3, and 12.7 $\mu$m
(790, 890, 1160, 1300, 1610, and 3040 cm$^{-1}$),
dominate the mid-infrared emission of our own
and other galaxies.
Figure 2 shows the 600 -- 2000 cm$^{-1}$ spectrum
of one region within our own Galaxy, NGC 7023 [2].
This figure shows most of the IEFs.
Note that this figure is plotted versus cm$^{-1}$, to be helpful to
chemists, but that the astronomical convention is to plot
spectra versus $\mu$m and that the astronomical convention
will be followed for the remainder of this paper.

The IEFs, which have widths broader than natural
broadening mechanisms for lines in atomic gas or simple molecular
gas, were mainly first discovered by astronomical
observations in the 1970's [3 -- 6],
with the 12.7 $\mu$m IEF discovered only after higher spectral
resolution observations [7] were able to distinguish it from
Ne$^+$ emission at 12.8 $\mu$m (780 cm$^{-1}$).
The IEFs have been referred to by a variety of names in the literature,
with ``feature'' and ``band'' used interchangeably:
the ``unidentified emission features,''
the ``unidentified infrared bands,''
the ``infrared emission features,''
the ``aromatic infrared bands,''
and the ``polycyclic aromatic hydrocarbon (PAH) bands.''
The origin of the last two names will become apparent in later
paragraphs.

The IEFs are believed to arise in either interstellar grains or
interstellar molecules, and because their emission can account for
up to 30 -- 40\% of the overall emission of galaxies,
solving problems related to their composition, size, and origin
are essential for understanding the overall energy budget of radiation
from galaxies.
Within our galaxy, the IEFs are observed in a wide variety of environments.
One important clue to the composition of the carrier of the IEFs, however,
is the astronomical observation that
the strength of the 7.7 $\mu$m IEF shows a strong correlation 
with the ratio of carbon to oxygen in planetary nebulae
[8 -- 10].
Ordinarily there is more oxygen than carbon in interstellar gas and in stars,
which results in an interstellar chemistry
in which common elements most often
combine with hydrogen and oxygen to form molecules and grains.
The heavier elements in such an oxygen-rich environment generally condense
into solid dust grains composed of oxides of these elements, such 
as silicate grains which are commonly observed in the ISM.
In some dying stars, however, convection within the star 
dredges up so much carbon, produced in nuclear reactions at the star's center,
that the composition of the outflowing gas from a dying star changes to a
peculiar situation in which the gas has more carbon than oxygen.
This carbon-rich gas alters the chemistry of grain and molecule formation
in the outflowing gas, preferentially forming molecules and grains
which are rich in hydrogen and carbon.
The observed correlation between the 7.7 $\mu$m IEF 
strength and the C/O ratio in planetary nebulae
strongly suggests that the IEFs form from dust and/or molecules condensed 
in the outflows from dying stars with carbon-rich outflows.
Once the IEFs, like other grains and molecules, are formed,
the gas ejected from the dying star slowly mixes with the
surrounding gas and dust within the galaxy and becomes part of the general
ISM.  
The ISM in galaxies, therefore, 
contains a mixture of molecules and
grains formed in oxygen-rich environments and carbon-rich environments.

Astronomical observations show strong emission from the IEFs in many 
regions of star formation, where dust and gas are interacting
with the ultraviolet (UV) and visible radiation emanating from young 
stars or from stars in the process of forming (protostars). 
Two examples of star formation regions with strong IEF emission are
reflection nebulae, where grains within a cloud of gas and dust scatters
the light of a nearby young star, and 
photodissociation regions surrounding 
H II regions (see reviews by [11, 12]).
An H II region is the astronomical name for a cloud of ionized hydrogen gas
surrounding a young star hot enough to ionize hydrogen;
a photodissociation region is the astronomical name for the surface
of a cloud of molecular hydrogen gas, where the UV radiation from a hot star
is in the process of photodissociating the molecular hydrogen.
Regions of the ISM are commonly referred to by the
state of their hydrogen gas -- neutral (H I), ionized (H II), or
molecular (H$_2$) -- because 71\% of the mass of
interstellar gas is composed of hydrogen.
The hydrogen gas in reflection nebulae can be either neutral or molecular,
but not ionized.

Astronomical mapping and images of the IEFs in the H II regions and adjacent
photodissociation regions in the star-forming Orion Nebula and the planetary
nebula NGC 7027 suggest that 
the IEFs are concentrated in a narrow zone between the ionization front
and the photodissociation front
[13, 14].
The ionization front is where ionized hydrogen abruptly
changes to neutral hydrogen
at a distance from the ionizing star where 
all of the ionizing photons from the star has been absorbed by
intervening hydrogen gas and re-radiated as several lower energy,
non-ionizing photons.
The ionization front
can be delineated by astronomical observations of recombination lines 
of neutral hydrogen; such lines are strong in ionized hydrogen but
absent from neutral hydrogen.
The photodissociation front can be traced by astronomical observations
of H$_2$ emission; the same UV photons which can dissociate H$_2$ have an
alternate path in which they cascade through the electronic levels of
H$_2$ to populate vibrational-rotational levels of H$_2$, which are then
observed as H$_2$ emission lines in the near-infrared (1 -- 2.5 $\mu$m
or 4000 -- 10,000 cm$^{-1}$). 

The IEFs are also observed in regions of low UV intensity, such as
the surfaces of molecular clouds and the general ISM
where the illumination source is the diffuse interstellar radiation field
[15 -- 17].
The ubiquity of IEF emission within our Galaxy, within star formation
regions, within carbon-rich planetary nebulae, and within the general
ISM,  makes understanding
their carriers an essential part of understanding 
how our Galaxy emits, absorbs, and reprocesses radiation.
The prevalence of IEF emission in star formation regions, and the
recent detection of IEF absorption toward protostars 
[18 -- 23],
also illuminates
the important role that the IEF carriers play in the material from
which stars and planets form.

The IEFs were first proposed to be due to aromatic hydrocarbons by
Duley \& Williams [24],
based on wavelength coincidences between the IEFs
and laboratory spectra of aromatic hydrocarbons.  
Another clue to the
composition of the IEF emitters 
was contributed by Sellgren et al. [25] and 
Sellgren [26], when we discovered 1 -- 5 $\mu$m (2000 -- 10,000 cm$^{-1}$)
continuum emission in 
reflection nebula that was
not due to scattered starlight.  
Starlight which is scattered from grains is strongly polarized,
and this infrared continuum emission in reflection nebulae
was unpolarized [27].
This continuum emission had a color 
temperature around 1000 K, and was spatially
associated with IEF emission in these sources.
The color temperature of the continuum emission showed no decrease with
distance from the exciting star, as would be expected for thermal emission
by dust in equilibrium with the star's radiation field.
Furthermore the stellar radiation field was far too weak to heat grains
to equilibrium temperatures of 1000 K at the observed nebular distances.
Temperatures of $\sim$ 30 -- 60 K were expected instead, 
and in fact dust in this temperature range 
had already been observed [28, 29] at far infrared wavelengths 
(40 -- 400 $\mu$m or 25 -- 250 cm$^{-1}$).
We proposed a new paradigm for interstellar dust,
in which the dust size distribution
contained a component of 1 nm sized grains which were so
small that a single UV photon 
absorbed from a nearby star could briefly heat the
grain to temperatures as high as 1000 K.  
Tiny particles of this size has been proposed earlier [30, 31].
These tiny grains spend the
vast majority of their time at low temperature, but when they undergo
stochastic heating they can for a microsecond or millisecond radiate
emission at high temperatures [25, 26]. 
This model explains the high color temperature
of the continuum emission, the constancy of the color temperature with
distance from the exciting star, and the lack of polarization.
Once we had proposed this model, several groups quickly realized the
importance of this emission mechanism to the newly emerging data
from the Infrared Astronomical Satellite (IRAS),
launched in 1984,
because the tiny grains, as they cooled radiatively, would be characterized
by progressively lower color temperatures 
with time and thus contribute significantly
to emission in the IRAS 12 and 25 $\mu$m (400 and 830 cm$^{-1}$) broad-band
photometric filters 
[32 -- 34].

The real breakthrough in understanding the IEFs, however, was made
by L\'eger \& Puget [35] and Allamandola et al. [36].
They were the first to combine the aromatic hydrocarbon idea 
[24] with a 1 nm particle size [26], 
and to propose that the IEFs were due to polycyclic
aromatic hydrocarbon (PAH) molecules with a size of roughly 1 nm or smaller.
Other aromatic materials have been proposed for the IEFs,
including tiny
grains composed of hydrogenated amorphous carbon, quenched carbonaceous
composite, and coal-like materials 
[37 -- 42].
Virtually all astronomers working on the IEFs agree on the
aromatic nature of the IEFs.
There is still debate over the exact nature of
the interstellar aromatics, but many favor the PAH hypothesis
because it so neatly ties together well-studied aromatic materials
with the 1 nm sizes suggested by the continuum emission associated
with the IEFs.

The goal of the astronomical observational research 
of myself and my collaborators
has been to test specific
predictions of the PAH hypothesis, and to urge chemical theorists
and laboratory astrochemists to make modifications in the PAH hypothesis
when the predictions and observations do not match.  This interplay 
between the observations, theory, and laboratory work has proven very
fruitful in refining our understanding of the size, composition, 
ionization state, and hydrogenation of the material responsible for the 
IEFs in the ISM.

\subsection{A Challenge: Approximating Interstellar Conditions in the
Laboratory}

One of the most difficult challenges in matching astronomical observations
and laboratory measurements of candidate materials for the IEFs is the
vast gulf between interstellar conditions and laboratory conditions.
The densities in interstellar space, even in the densest interstellar
clouds, are so low that they exceed the capabilities of the 
best vacuum systems on Earth.
The vast majority of published laboratory work has been measurements
of solid-phase, room temperature aromatic materials 
in absorption, which are then compared to astrophysical observations of
$\sim$1000 K aromatic materials in emission in extremely low density 
environments.
Some progress has been made in approximating the low densities of
interstellar conditions by use of inert-gas matrix isolation techniques,
but even in these cases matrix interactions introduce
frequency shifts in the wavelengths of absorption lines of unknown
magnitude and direction.
Other progress has been made in measuring aromatic molecules
in emission in the laboratory, which has helped to quantify the
temperature dependence of wavelength shifts in PAHs.
Recent evidence, discussed below, suggests that the PAHs that have
been best-studied in the laboratory are too small to approximate
interstellar PAHs.
Furthermore, astrophysical theoretical calculations predict that PAHs
should be ionized in the best-observed regions of the ISM. 
Both chemical quantum calculations and laboratory measurements demonstrate
marked changes in the relative intensities and central wavelengths
of PAH bands with ionization state.
All of these mismatches between interstellar conditions and
laboratory conditions make it difficult to achieve a firm identification
of the material(s) responsible for the IEFs.

\section {The IEFs and the Hardness of the UV Radiation Field}

One of the early predictions of the PAH model is that PAH molecules should
only be excited by UV radiation, since laboratory absorption curves of 
small, neutral PAHs
show a sharp cutoff in their absorption at UV wavelengths with little or no
absorption at visible wavelengths. 
Sellgren et al. [43] tested this prediction by using observations from
the IRAS satellite to search for infrared emission in reflection nebulae
excited by stars of different effective temperature, $T_{\rm eff}$.
We first verified that each reflection nebula was indeed excited by the
central star of the nebula, by requiring that the dust temperature
derived from the the ratio of 60 $\mu$m (170 cm$^{-1}$)
intensity to 100 $\mu$m (100 cm$^{-1}$) intensity 
reach a peak value at
the star.  The emission at 60 and 100 $\mu$m
is thought to primarily come from grains, with a size around 100 nm, 
which are in
equilibrium with the radiation field, and whose temperature therefore will
increase with proximity to the heating star.  We then examined the 12 $\mu$m
emission of each reflection nebula, at a nebular position offset from the
star so that stellar emission would not contaminate the 12 $\mu$m data,
and compared it to the total infrared emission at the same spatial location.
For each nebula we then measured $R(12/{\rm total})$, the ratio of the total
flux in the IRAS 12 $\mu$m band to the total infrared flux.  
The IRAS 12 $\mu$m band we assumed to be dominated by IEF emission and its
associated continuum emission, as
had been shown by mid-infrared spectroscopy for some of the sources in
our sample [34].
We were expecting $R(12/{\rm total})$ to show 
a precipitous drop for cool stars ($T_{\rm eff}$ $<$ 10,000 K),
where the fraction of total stellar radiation radiated in the UV was small.
Much to our surprise, the value of 
$R(12/{\rm total})$ was independent of $T_{\rm eff}$,
over the range $T_{\rm eff}$ = 5,000 K -- 22,000 K,
as illustrated in Figure 3.  
Our observational results required that the material responsible for
emitting the IRAS 12 $\mu$m emission in reflection nebulae absorb not
only at UV but also at visible wavelengths.
This discovery altered the astronomical community's perception of PAHs
as being small in size and neutrally charged, and drove theorists
and laboratory
astrochemists to consider both larger PAHs and ionized PAHs in their
models and laboratory work.
This is because increasing the PAH size and ionizing PAHs both have
the effect of extending the absorption cross-section of PAHs out to
visible wavelengths.

The 12 $\mu$m IRAS filter bandpass is wide and encompasses several IEFs
as well as their associated continuum.  The material responsible
for the continuum has never been identified, and the possibility
exists that it is not due to the same material that produces
the IEFs.  In this case, it might be possible that the IRAS 12 $\mu$m
emission we observed in reflection nebulae illuminated by cool stars
is due to continuum emission alone without any contribution from
the IEFs. Thus, when the Infrared Space Observatory (ISO) was launched
in 1995, we
embarked on an observational program to obtain spectra of
some of the reflection nebulae studied by Sellgren et al. [43],
particularly reflection nebulae illuminated by both cool and hot
stars which had similar values of $R(12/{\rm total})$ .
These observations have been made with ISO's mid-infrared camera (ISOCAM) 
combined with its circular variable filter (CVF), which allow us to
obtain low-resolution ($\lambda / \Delta \lambda $ = 40) spectra,
at 5 -- 15 $\mu$m (600 -- 2000 cm$^{-1}$), 
simultaneously at roughly a thousand spatial
positions across each nebula.
The ISO spectra of these reflection nebulae, therefore, should unambiguously
determine whether the IEFs are present in these sources.

The first ISO results from Uchida et al. [44] was our discovery
that IEFs are clearly
detected in vdB 133, a reflection nebula illuminated by a binary system with
very little UV radiation.  The binary system, a luminous
star with $T_{\rm eff}$ = 6,800 K plus a fainter star with 
$T_{\rm eff}$ = 12,000 K,
provides a ratio of UV ($\lambda$ $<$ 400 nm) to total stellar flux which is a
factor of four lower than more typical reflection nebulae 
which are illuminated by stars with $T_{\rm eff}$ $\sim$ 20,000 K.
Yet, despite the softer radiation environment, the IEF spectrum in this
UV-poor environment is
very similar to IEF spectra observed in sources with much harsher UV
environments, as shown in Figure 4.

\subsection{New Results: Laboratory Analogs and the
IRAS Data on R(12/total)}

I am currently collaborating with J. Pizagno, K. Uchida, and M. Werner
on a comparison of various laboratory
and theoretical candidates for the IEF carriers with the IRAS
observations [43] of $R(12/{\rm total})$,
to quantify which materials can reproduce the lack of dependence
of $R(12/{\rm total})$ on $T_{\rm eff}$
(Pizagno et al. 2000, in preparation).
We are convolving the UV and visible 
laboratory and theoretical absorption curves of
different materials with the energy distributions
emitted by stars of different
$T_{\rm eff}$, and comparing the results to our IRAS observations.
Figure 3 shows one sample result of our calculations, where we
compare the astronomical observations to the predicted results for
two small PAH molecules, neutral naphthalene [45] and singly ionized
naphthalene (from F. Salama 1999, private communication).
These new results show that small neutral PAHs are completely unable to
provide enough visible absorption to explain the IEF emission observed
around UV-poor sources.
They also show that while ionized PAHs have more visible absorption
than their neutral counterparts, small PAHs even when ionized also
cannot explain the IRAS observations.

\subsection{Need for Data and Theory for Larger PAHs and Varying Ionization}

It seems likely that a combination of ionization and larger PAH size
will be required to explain the astronomical observations.
We strongly encourage laboratory astrochemists to measure 
parameters for larger PAHs, both neutral and ionized.
We also strongly encourage theoretical astrochemists to calculate
parameters for larger PAHs, both neutral and ionized, particularly
for PAH sizes and/or ionization states that are not currently
accessible by laboratory techniques.

\section {PAH Ionization}

Uchida et al. [46]
have just completed a more comprehensive ISO study of IEFs in reflection
nebulae, in which we have made unexpected
observations which have important ramifications for models of
PAH ionization.
PAH ionization models predict that PAHs should be primarily positively
charged in regions of high UV radiation, such as reflection nebulae,
and primarily neutral or negatively charged in regions of low UV
radiation, such as the diffuse ISM and cirrus clouds
[47 -- 50].
The UV radiation intensity is generally characterized by the variable
$G_0$, where $G_0$ = 1 corresponds to the UV radiation field in the
solar neighborhood.
Laboratory and theoretical studies show that
the ratio of PAH emission at 6 -- 10 $\mu$m 
(1000 -- 1700 cm$^{-1}$)
to the PAH emission at 10 -- 14 $\mu$m 
(700 -- 1000 cm$^{-1}$)
is a sensitive function of PAH
ionization state, with this ratio varying by at least a factor of 10
between ionized and neutral PAHs 
[51 -- 59].

We have measured the ratio of IEF emission at 6 -- 10 $\mu$m to
the IEF emission at 10 -- 14 $\mu$m, observed in reflection nebulae
with $G_0$ varying from 20 to 6$\times$10$^4$. 
Figure 5 shows that we
see little or no evidence for any change in this ratio with 
$G_0$  (and thus PAH ionization state).
This provides a major puzzle for the PAH hypothesis.

\subsection{PAH Ionization: Astrochemical Theory and Laboratory Data Needed}

There are several ways to resolve
the discrepancy between the observations and theory
for the predicted dependence of the PAH ionization state on
the UV field.
The calculated ionization states of PAHs are somewhat sensitive to
the assumed PAH size.  
The amount by which 
the ratio of fluxes at 6 -- 10 $\mu$m to 10 -- 14 $\mu$m
changes also depends on PAH size, with smaller changes in the ratio
for larger PAHs.
The calculated ionization states of PAHs are also very sensitive to
the uncertain PAH recombination rate, which requires better laboratory
measurements and/or better theoretical predictions for the recombination
rates, particularly as a function of PAH size.
These are all areas in which astrochemistry in the laboratory and
theoretical calculations can really make a difference in interpreting
the astronomical observations.

\section {The Full-Width at Half Maximum of the 7.7 $\mu$m IEF}

Another unexpected discovery Uchida et al.
[46] have made with our recent ISO observations
concerns spectral changes in the
IEFs at low levels of illumination.
We have quantitatively compared the IEF spectra (Fig. 4) of three sources we
observed ourselves, vdB 133, vdB 17 (NGC 1333), and vdB 59 (NGC 2068),
to the IEF spectra of published spectra of NGC 7023 ([2]; Fig. 1)
and $\rho$ Oph [15]. 
We have found no evidence for any systematic spectral differences with 
$T_{\rm eff}$.
Our new ISO observations, however, find 
that the full width at half maximum (FWHM) of
the 7.7 $\mu$m IEF is dependent on the distance between star and nebula in
vdB 17 [46].
Figures 6 and 7 show that
in the most distant regions
of vdB 17, corresponding to the lowest UV
illumination levels, $G_0$ = 20 -- 60,
the 7.7 $\mu$m IEF becomes significantly broader and begins to
blend with the 8.6 $\mu$m IEF.
This effect has now been observed in a second reflection nebula,
Ced 201 [60], although they do not give the $G_0$ values
at which the broadening of the 7.7 $\mu$m IEF becomes significant.
None of the other IEFs appear to change width with stellar distance
in either source.
We also find no broadening of the 7.7 $\mu$m IEF or other IEFs
in vdB 59, down to $G_0$ = 200.
More significantly, the spectrum of $\rho$ Oph [15]
at $G_0$ = 40 shows no sign of a broader 7.7 $\mu$m IEF or any
blending of the 7.7 $\mu$m IEF with the 8.6 $\mu$m.

The unexpected broadening of the 7.7 $\mu$m IEF at low UV illumination
levels, in some sources but not others, is very much a mystery.
One possibility is that the intermittent broadening
of the 7.7 $\mu$m feature depends on two parameters, one being $G_0$,
for instance, and the other being some factor such as $T_{\rm eff}$.
Stellar $T_{\rm eff}$ seems like a good starting point for investigation,
because vdB 17 and Ced 201, which are illuminated by stars with
$T_{\rm eff}$ = 10,000 -- 11,000 K,
show the broadening of the 7.7 $\mu$m IEF at low $G_0$ values, while
$\rho$ Oph, which shows no such broadening at $G_0$ = 40, is illuminated
by a pair of stars with $T_{\rm eff}$ = 22,000 K.  

Two other ideas seem like promising avenues to explore.
One is the recent ISO observations of Moutou et al. [61], illustrated in
Figure 8. 
We find that the 7.7 $\mu$m IEF, when
observed at higher spectral resolution ($\lambda / \Delta \lambda$ = 1800),
appears to break up into at least three different features [61],
at 7.45, 7.6, and 7.8 $\mu$m (1280, 1320, and 1340 cm$^{-1}$).
Previous astronomical spectra 
at $\lambda / \Delta \lambda$ = 100 -- 200
[62, 63] resolved the 7.7 $\mu$m IEF
into two features at 7.6 and 7.8 $\mu$m.
Other astronomical observations [9, 10]
at a lower resolution of $\lambda / \Delta \lambda$ = 50, where
only the central wavelength of the 7.7 $\mu$m IEF could be measured,
demonstrated that the 7.7 $\mu$m IEF
central wavelength depends on the physical
conditions in the emitting region (whether or not the hydrogen gas 
is ionized, for instance).
Since the results of [9, 10] almost surely result
from subtle changes in the relative strengths of the subcomponents
of the 7.7 $\mu$m IEF,
then the broadening of the 7.7 $\mu$m IEF observed at low resolution
[46] could reflect more obvious changes in the 
relative strengths
of the different subcomponents observed at higher resolution [61 -- 63].

The other idea deserving further exploration
is related to the fundamental reason why the IEFs are
so much broader than gas molecules or grains moving at the Doppler speeds
(1 km s$^{-1}$ or less) typical of these interstellar regions.
There is no clear consensus on whether the IEF widths
are due to physical processes (such as timescales for internal
molecular processes, which depend on the size and temperature of the molecule)
or to mixtures of aromatics with 
slightly different compositions, sizes, and
central wavelengths.
Observations of the IEFs in absorption suggest that the mixture of
IEF carriers may be dominant in determining the central wavelength
and width of the IEF emission, as discussed in the next section,
and if this is the case, then the astronomical observations of
changes in the central wavelength or FWHM of the 7.7 $\mu$m IEF
could reflect changes in the size distribution or composition of interstellar
aromatics under different physical conditions.

\subsection{The 7.7 Micron IEF Width: Need for Laboratory and Theoretical Work}

Astrochemical work is desperately needed to formulate some theoretical
framework for why the 7.7 $\mu$m width changes, so that astronomical
observations can be made to test this framework.
As described above, the intermittent broadening
of the 7.7 $\mu$m IEF could depend on two parameters, one being $G_0$,
for instance, and the other being some factor such as stellar temperature.
Can laboratory or theoretical work provide any insights into why
these two parameters, or some other set of parameters, might affect
the 7.7 $\mu$m width?
Does the broadening of the 7.7 $\mu$m IEF observed at low resolution
reflect changes in the
relative strengths
of the different subcomponents at 7.45, 7.6, and 7.8 $\mu$m observed 
at higher resolution [61]?
Or do the relative strengths of the different subcomponents
of the 7.7 $\mu$m IEF stay the same, but the width of each
subcomponent change?
What physical change would be needed in the IEF carriers for
the width to change?
Could it be that the size distribution
or composition of interstellar aromatics change with different
physical conditions?
Or is it some other phenomenon entirely?
The answer to the questions posed by observations of the 7.7 $\mu$m
IEF are rooted in the ongoing debate as to the origin of the
IEF widths themselves.

\section {Aromatics in Absorption}

Until recently, all astronomical observations of aromatic hydrocarbons
in the ISM were of the IEFs in emission.  The absorption signatures of the IEFs
had not been seen.  We have begun a program to try to trace the
abundance and chemical changes of aromatic hydrocarbons during their
residence time in molecular clouds.  A broad 3.25 $\mu$m (3080 cm$^{-1}$)
absorption feature along the line-of-sight to the 
protostar Mon R2/IRS-3 was tentatively detected,
and its discovery confirmed with better
spectra, by Sellgren et al. [18, 19].  We proposed that this feature
is due to the C--H stretch absorption of cold aromatic hydrocarbons
within the molecular cloud. 
Laboratory work on PAHs [64] has shown that the
central wavelength of PAH features shifts to increasing wavelength
with increasing temperature.
This predicts that aromatics in absorption should be blue-shifted by about
the observed amount, relative to aromatics in emission,
because aromatics in emission are emitting at temperatures
of $\sim$1000 K, while aromatics in absorption are
absorbing at much lower temperatures, $\sim$ 10 -- 80 K.
In Brooke et al. [21, 22] we have now detected the 3.25 $\mu$m absorption
feature toward a total of five 
protostars embedded in molecular clouds.

\subsection {New Results: Comparing Aromatics in Emission and Absorption}

The current observations of aromatics in absorption, compared
to aromatics in emission, at first glance seem to have a simple
interpretation.
But there is an interesting complication.  Ground-based [65]
and ISO [66] spectra of
sources near the Galactic Center reveal a narrow absorption feature
near 3.28 $\mu$m (3050 cm$^{-1}$), 
which these authors identify with aromatic
hydrocarbons.  
If this feature is due to the 3.29 $\mu$m (3040 cm$^{-1}$) 
IEF emitter, observed in absorption
through the diffuse ISM in our Galaxy, it should 
spend the vast majority of its time at cold temperatures and 
thus have a wavelength blue-shifted relative to the 3.29 $\mu$m IEF
in emission, similar
to the blue-shifted wavelength (3.25 $\mu$m; 3080 cm$^{-1}$)
observed in absorption toward molecular clouds.
The longer wavelength of the 3.28 $\mu$m absorption suggests either
that the aromatics are at a higher temperature, perhaps due to localized
heating by sources in the Galactic Center, or that the composition, 
sizes, or ionization state of the aromatics 
observed in the diffuse ISM are different from
the aromatics observed in molecular clouds.

The differences in observed width are also confusing.
Laboratory studies [64] have shown
that as individual PAHs are heated, 
their central
wavelengths both shift to longer wavelengths and the width of the PAH
band broadens.
Figure 9, which has not been published elsewhere, illustrates
the contrast between the short wavelength and broad width of the 
3.25 $\mu$m aromatic 
absorption in molecular clouds [19], and the longer wavelength and narrow
width of the 
3.28 $\mu$m aromatic absorption towards the Galactic Center [66]. 
Both spectra have $\lambda / \Delta \lambda$ = 1000.
Figure 9 also shows a spectrum of the 3.29 $\mu$m IEF in emission
in the reflection nebula NGC 7023, from the work in progress of
myself, C. Moutou, A. L\'eger, L. Verstraete, M. Werner, M. Giard,
and D. Rouan.
This spectrum, with $\lambda / \Delta \lambda$ = 1000, demonstrates
that the observed width of 
the 3.29 $\mu$m IEF in emission (Sellgren et al. 2000, in preparation) 
is similar to or narrower than
the 3.25 $\mu$m feature in absorption [19],
again in contradiction to the temperature model.
This suggests
that the observed wavelength shifts and widths are not due to temperature
effects but rather to composition, size distribution, ionization,
or other effects.

This mystery is underscored by recent observations of
an absorption feature at 6.18 $\mu$m (1620 cm$^{-1}$), 
believed to be the
aromatic C--C stretch, toward five
hot, mass-losing stars and two Galactic Center sources in ISO spectra 
[67].  A possible absorption feature at 6.24 $\mu$m
(1600 cm$^{-1}$)
toward several protostars has also been tentatively identified
as the C--C stretching mode of aromatic hydrocarbons 
[20, 23].
If the 6.18 $\mu$m absorption toward the sources observed by 
Schutte et al. [67] arises from aromatics in the diffuse ISM,
the aromatics should again be cold with blue-shifted, narrow absorption
features.
The 6.24 $\mu$m absorption feature observed by Keane et al. [23] 
toward protostars in
molecular clouds, if due to aromatics, should also arise from
cold aromatics with blue-shifted, narrow absorption
features.
The astronomical observations, however, do not fit this picture.
The 6.18 $\mu$m absorption feature observed through the diffuse ISM [67]
is blue-shifted relative to the 6.22 $\mu$m (1610 cm$^{-1}$)
IEF in emission [68],
as expected, but the 6.24 $\mu$m absorption in molecular clouds is
observed at a similar or slightly longer
wavelength as the 6.22 $\mu$m IEF in emission.
Furthermore, the 6.18 $\mu$m absorption feature observed through the
diffuse ISM is significantly narrower than the 6.24 $\mu$m absorption
feature observed towards molecular clouds, as shown in
Figure 10 of Keane et al. [23].
Again this is strong evidence that both the central wavelengths and
widths of aromatics observed in absorption and emission in the
ISM are determined not by temperature or molecular physics of a single
molecule, but rather by the composition mixture, size distribution,
or mix of ionization states
of aromatics observed along a particular line-of-sight.

\subsection{Need for Laboratory Work on PAHs in Water Ice Matrices}

One possible way to reconcile the temperature picture of the wavelength
and width of aromatic absorption and emission with the astronomical
observations is to consider whether aromatics in molecular clouds
might be embedded in ice mantles on grains.
Solid H$_2$O ice is observed toward all sources in molecular clouds
with detected 3.3 or 6.2 $\mu$m aromatic absorption features.
New astrochemical laboratory measurements 
at NASA's Ames Research Center are currently underway
to measure the effect of a surrounding ice matrix on PAH absorption
(M. Bernstein 2000, private communication).
These results, once published, may clarify our understanding of aromatic
absorption in molecular clouds.
Measurements are needed for both neutral and ionized PAHs.

\section {Interstellar ``Diamond-like'' Carbon}
 
The broad 3.47 $\mu$m (2880 cm$^{-1}$)
absorption feature (FWHM $\approx$ 0.1 $\mu$m or 80 cm$^{-1}$) 
was first
noted in ground-based spectra of four protostars by 
Allamandola et al. [69]. They suggested that the 
feature might be due to the C--H
stretch absorption of solo hydrogens attached to $sp^3$ bonded carbon
clusters, the ``diamond''-like form of carbon.  The feature was
present in every molecular cloud source looked at by Brooke et al. [21].
Interstellar microdiamonds, formed outside our solar system,
have been identified in meteorites 
[70, 71]
and so must therefore exist and survive in the diffuse ISM.
Our observations [21, 22] of the 3.47 $\mu$m absorption feature,
however, reached the startling conclusion that the optical depth of 
the 3.47 $\mu$m absorption feature is not correlated with
the depth of silicate absorption, as might be expected for
two refractory minerals such as diamonds and silicates
(Figure 10).
Instead, we found that the 3.47 $\mu$m
optical depth is strongly correlated
with the depth of the H$_2$O ice band, a very volatile
ice that cannot exist outside molecular clouds
(Figure 11).

\subsection {Need for Laboratory or Theoretical Frequencies for 
C--H Bonds on the Surfaces of Microdiamonds}

Under the assumption that the 3.47 $\mu$m feature is due to 
C--H bonds, Brooke et al. [21] interpreted the correlation
with H$_2$O ice as indicating that both C--H bonds and H$_2$O ice form
in step on molecular cloud dust by hydrogen addition reactions.
However, we noted that other identifications of the 
3.47 $\mu$m feature are also possible.  
One fertile region for astrochemical research is to better establish
the formation mechanism and absorption spectrum of interstellar
``diamond''-like carbon of interstellar size 
(1 nm; [70]) with surface hydrogen atoms.
Predictions or measurements of the complete vibrational spectrum,
including additional modes which could be searched
for in astronomical spectra, would be particularly useful.
It would also be useful to search for alternative identifications
for the 3.47 $\mu$m feature
with simple molecules, composed of cosmically abundant elements
(Fig. 1), which would have sublimation temperatures close to
that of H$_2$O ice.
Again, predictions or measurements of other vibrational modes
that could be used to test the identification would be helpful.

\section {C$_{60}^+$ in the interstellar medium}

The fullerene C$_{60}$ was first discovered in the laboratory by 
Kroto et al. [72].  They proposed that this novel form of 
aromatic carbon could potentially
play an important role in the ISM.  Theoretical studies
of dust formation in carbon-rich stellar mass-loss 
[73 -- 75] suggested that fullerenes
could be formed in such carbon-rich environments and expelled into the
ISM.  Laboratory studies have also shown that one product
of the photoerosion of hydrogenated amorphous carbon grains is fullerenes
[76].  The existence of 
fullerenes such as C$_{60}$ and C$_{70}$
in our solar system, for instance in meteoritic samples, is hotly debated
[77, 78]. 

Searches for C$_{60}$ in the ISM, 
through its UV absorption band
at 386 nm, have placed stringent limits of $<$ 0.01\% of the cosmic carbon
abundance in C$_{60}$ [79, 80].
The dominant ionization state of C$_{60}$ in the ISM, 
however, is predicted to be C$_{60}^+$
[47, 49, 50].
Foing \& Ehrenfreund [81, 82] detected diffuse interstellar
bands (see [83] for a review of the diffuse interstellar bands)
at 958 and 963 nm, which they argued were due to C$_{60}^+$, based on a
comparison to laboratory data [84].

This inspired Moutou et al. [61]
to use ISO data to search for the 7.1 and 7.5 $\mu$m
(1331 and 1406 cm$^{-1}$)
vibrational emission lines of C$_{60}^+$ in NGC 7023, a reflection
nebula with strong IEF emission.
Such a search requires high signal-to-noise and high spectral
resolution ($\lambda / \Delta \lambda$ = 1800), 
because these features, if present, would appear
as weak bumps on the blue wing of the strong aromatic 7.7 $\mu$m 
IEF.  From our ISO spectra (Figure 8), we 
place an upper limit on C$_{60}^+$ in NGC 7023
of $<$ 0.3\% of interstellar carbon [61].

The central star of NGC 7023,
HD 200775, is known to have abnormally
weak diffuse interstellar bands [85].
If the diffuse interstellar bands observed at 958 and 963 nm are
due to C$_{60}^+$, then the general weakness of the diffuse interstellar bands
in NGC 7023 might naturally lead to a failure to detect C$_{60}^+$ in the
mid-infrared toward this source.  Astronomers can place more robust limits on
the abundance of fullerenes in the ISM by searching
for the 7.1 and 7.5 $\mu$m C$_{60}^+$ lines towards a larger number of lines
of sight, particularly those in which the diffuse interstellar bands
are strong.

\subsection {Laboratory and Theoretical Work Needed for Fullerenes}

Laboratory astrochemists can also contribute greatly to the search for
interstellar fullerenes by further experiments.
The assignment of 
the interstellar bands at 958 and 963 nm 
to C$_{60}^+$
was based on a comparison of gas-phase C$_{60}^+$ in the ISM 
with C$_{60}^+$ measured in inert-gas matrices, with the
effects of matrix shifts on the wavelengths being unknown.
The 7.1 and 7.5 $\mu$m C$_{60}^+$ bands were also measured in
inert-gas matrices, again introducing an unknown amount of matrix
shift to the wavelengths.
Laboratory measurements of all these bands in the gas phase would 
help immensely to determine whether detections of astronomical
absorption features near
the laboratory wavelengths of C$_{60}^+$ prove or disprove the
existence of C$_{60}^+$ in the ISM.
Finally, by symmetry, there should be two longer wavelength vibrational
modes of C$_{60}^+$ which were not measured by Fulara et al. [84].
Laboratory wavelengths for these two additional modes would open up
new avenues for astronomical observations of C$_{60}^+$.

\section{Conclusions}

I have emphasized in this paper a number of puzzling astronomical observations
of aromatics in the ISM, and related molecules, where understanding 
of these problems would
greatly benefit from new experiments in laboratory astrochemistry
and progress in chemical theory.
The study of the ISM is an area where the interaction of astronomy,
physics, and chemistry has proven particular fruitful in reaching
new conclusions about the universe that surrounds us, and I strongly
encourage astrochemistry groups to tackle these experimental and
theoretical problems as part of this on-going interaction.

\acknowledgments

I would like to thank Tim Brooke for contributions to the text
on the 3.47 $\mu$m feature. I am grateful to Max Bernstein,
Jean Chiar and Farid Salama for communicating
data in advance of publication, and to Jacquie Keane for sharing a preprint
before the refereeing process was complete.
I also thank James Pizagno for speedy assistance with Figure 3.

\section{References}

\hang\noindent
[1] A.N. Cox, ed.
2000, Allen's Astrophysical Quantities, (4th ed.;
New York: AIP/Springer), 29.

\hang\noindent
[2] D. Cesarsky, J. Lequeux, A. Abergel, M. P\'erault, E. Palazzi, 
S. Madden, D. Tran, 
Astron. Astrophys. 315 (1996) L305.

\hang\noindent
[3] F.C. Gillett, W.J. Forrest, K.M. Merrill,
Astrophys. J. 183 (1973) 87.

\hang\noindent
[4] M. Cohen, C.M. Anderson, A. Cowley, G.V. Coyne, W. Fawley, T.R. Gull, 
E.A. Harlan, G.H. Herbig, F. Holden, H.S. Hudson, R.O. Jakoubek, 
H.M. Johnson, K.M. Merrill, F.H. Schiffer, B.T. Soifer, B. Zuckerman, 
Astrophys. J. 196 (1975) 179.

\hang\noindent
[5] K.M. Merrill, B.T. Soifer, R.W. Russell,
Astrophys. J. 200 (1975) L37.

\hang\noindent
\hang\noindent
[6] R.W. Russell, B.T. Soifer, S.P. Willner,
Astrophys. J. 217 (1977) L153.

\hang\noindent
[7] P.F. Roche, D.K. Aitken, C.H. Smith, 
Mon. Not. Roy. Astron. Soc. 236 (1989) 485.

\hang\noindent
[8] P.F. Roche, D.K. Aitken, 
Mon. Not. Roy. Astron. Soc. 221 (1986) 63.

\hang\noindent
[9] M. Cohen, L. Allamandola, A.G.G.M. Tielens, J. Bregman, 
J.P. Simpson, F.C. Witteborn, D. Wooden, D. Rank, 
Astrophys. J. 302 (1986) 737.

\hang\noindent
[10] M. Cohen, A.G.G.M. Tielens, J. Bregman, F.C. Witteborn, 
D.M. Rank, L.J. Allamandola, D. Wooden, M. Jourdain de Muizon, 
Astrophys. J. 341 (1989) 246.

\hang\noindent
[11] J.L. Puget, A. L\'eger, 
Ann. Rev. Astron. Astrophys. 27 (1989) 161.

\hang\noindent
[12] L.J. Allamandola, A.G.G.M. Tielens, J.R. Barker, 
Astrophys. J. Suppl. 71 (1989) 733.

\hang\noindent 
[13] K. Sellgren, A.T. Tokunaga, Y. Nakada, 
Astrophys. J. 349 (1990) 120.

\hang\noindent
[14] J.R. Graham, E. Serabyn, T.M. Herbst, K. Matthews, 
G. Neugebauer, B.T. Soifer, T.D. Wilson, S. Beckwith, 
Astron. J. 105 (1993) 250.

\hang\noindent
[15] F. Boulanger, W.T. Reach, A. Abergel, J.P. Bernard, C.J. Cesarsky,
D. Cesarsky, F.X. D\'esert, E. Falgarone, J. Lequeux, L. Metcalfe,
M. P\'erault, J.L. Puget, D. Rouan, M. Sauvage, D. Tran, L. Vigroux,
Astron. Astrophys. 315 (1996) L325.

\hang\noindent
[16] K. Mattila, D. Lemke, L.K. Haikala, R.J. Laureijs, 
A. L\'eger, K. Lehtinen, C. Leinert, P.G. Mezger, 
Astron. Astrophys. 315 (1996) L353.

\hang\noindent
[17] T. Onaka, I. Yamamura, T. Tanabe, T.L. Roellig, L. Yuen, 
Pub. Astron. Soc. Japan 48 (1996) L59.

\hang\noindent
[18] K. Sellgren, R.G. Smith, T.Y. Brooke, 
Astrophys. J. 433 (1994) 179.

\hang\noindent 
[19] K. Sellgren, T.Y. Brooke, R.G. Smith, T.R. Geballe, 
Astrophys. J. 449 (1995) L69.

\hang\noindent
[20] W.A. Schutte, A.G.G.M. Tielens, D.C.B. Whittet, A. Boogert, 
P. Ehrenfreund, Th. de Graauw, T. Prusti, E.F. van Dishoeck, P. Wesselius,
Astron. Astrophys. 315 (1996) L333.

\hang\noindent
[21] T.Y. Brooke, K. Sellgren, R.G. Smith, 
Astrophys. J. 459 (1996) 209.

\hang\noindent
[22] T.Y. Brooke, K. Sellgren, T.R. Geballe, 
Astrophys. J. 517 (1999) 883.

\hang\noindent
[23] J.V. Keane, A.G.G.M. Tielens, 
A.C.A. Boogert, W.A. Schutte, D.C.B. Whittet, 
2000, Astron. Astrophys., in press.

\hang\noindent
[24] W.W. Duley, D.A. Williams, 
Mon. Not. Roy. Astron. Soc. 196 (1981) 269.

\hang\noindent 
[25] K. Sellgren, M.W. Werner, H.L. Dinerstein, 
Astrophys. J. 271 (1983) L13.

\hang\noindent 
[26] K. Sellgren, 
Astrophys. J. 277 (1984) 623.

\hang\noindent 
[27] K. Sellgren, M.W. Werner, H.L. Dinerstein, 
Astrophys. J. 400 (1992) 238.

\hang\noindent
[28] P.M. Harvey, H.A. Thronson, I. Gatley, 
Astrophys. J. 235 (1980) 894.

\hang\noindent
[29] S.E. Whitcomb, I. Gatley, R.H. Hildebrand, J. Keene, 
K. Sellgren, M.W. Werner, 
Astrophys. J. 246 (1981) 416.

\hang\noindent
[30] J.R. Platt, Astrophys. J. 123 (1956) 486.

\hang\noindent
[31] C.D. Andriesse, Astron. Astrophys. 66 (1978) 169.

\hang\noindent
[32] B.T. Draine, N. Anderson, 
Astrophys. J. 292 (1985) 494.

\hang\noindent
[33] J.L. Puget, A. L\'eger, F. Boulanger, 
Astron. Astrophys. 142 (1985) L19.

\hang\noindent 
[34] K. Sellgren, L.J. Allamandola, J.D. Bregman, 
M.W. Werner, D.H. Wooden, 
Astrophys. J. 299 (1985) 416.

\hang\noindent
[35] A. L\'eger, J.L. Puget, 
Astron. Astrophys. 137 (1984) L5.

\hang\noindent
[36] L.J. Allamandola, A.G.G.M. Tielens, J.R. Barker, 
Astrophys. J. 290 (1985) L25.

\hang\noindent
[37] A. Sakata, S. Wada, T. Tanabe, T. Onaka, 
Astrophys. J. 287 (1984) L51.

\hang\noindent
[38] A. Sakata, S. Wada, T. Onaka, A.T. Tokunaga, 
Astrophys. J. 320 (1987) L63.

\hang\noindent
[39] A. Borghesi, E. Bussoletti, L. Colangeli, 
Astrophys. J. 314 (1987) 422.

\hang\noindent
[40] A. Blanco, E. Bussoletti, L. Colangeli, 
Astrophys. J. 334 (1988) 875.

\hang\noindent
[41] W.W. Duley, 
Mon. Not. Roy. Astron. Soc. 234 (1988) 61P.

\hang\noindent
[42] R. Papoular, J. Conrad, M. Giuliano, J. Kister, G. Mille, 
Astron. Astrophys. 217 (1989) 204.

\hang\noindent
[43] K. Sellgren, L. Luan, M.W. Werner, 
Astrophys. J. 359 (1990) 384.

\hang\noindent
[44] K.I. Uchida, K. Sellgren, M. Werner, 
Astrophys. J. 493 (1998) L109.

\hang\noindent
[45] F. Salama, L.J. Allamandola, 
J. Chem. Phys. 94 (1991) 6964.

\hang\noindent
[46] K.I. Uchida, K. Sellgren, M.W. Werner, M.L. Houdashelt, 
Astrophys. J. 530 (2000) 817.

\hang\noindent
[47] E.L.O. Bakes, A.G.G.M. Tielens, 
Astrophys. J. 427 (1994) 822.

\hang\noindent
[48] E.L.O. Bakes, A.G.G.M. Tielens, 
Astrophys. J. 499 (1998) 258.

\hang\noindent
[49] F. Salama, E.L.O. Bakes, L.J. Allamandola, A.G.G.M. Tielens, 
Astrophys. J. 458 (1996) 621.

\hang\noindent
[50] E. Dartois, L. d'Hendecourt, 
Astron. Astrophys. 323 (1997) 534.

\hang\noindent
[51] D.J. de Frees, M.D. Miller, D. Talbi, F. Pauzat, Y. Ellinger, 
Astrophys. J. 408 (1993) 530.

\hang\noindent 
[52] J. Szczepanski, M. Vala, 
Astrophys. J. 414 (1993) 646.

\hang\noindent
[53] D.M. Hudgins, S.A. Sandford, L.J. Allamandola, 
J. Phys. Chem. 98 (1994) 4243.

\hang\noindent
[54] F. Pauzat, D. Talbi, Y. Ellinger, 
Astron. Astrophys. 293 (1995) 263.

\hang\noindent
[55] F. Pauzat, D. Talbi, Y. Ellinger, 
Astron. Astrophys. 319 (1997) 318.

\hang\noindent
[56] S.R. Langhoff, 
J. Phys. Chem. 100 (1996) 2819.

\hang\noindent
[57] D.J. Cook, R.J. Saykally, 
Astrophys. J. 493 (1998) 793.

\hang\noindent
[58] L.J. Allamandola, D.M. Hudgins, S.A. Sandford, 
Astrophys. J. 511 (1999) L115.

\hang\noindent
[59] D.M. Hudgins, L.J. Allamandola, 
Astrophys. J. 516 (1999) L41.

\hang\noindent
[60] D. Cesarsky, J. Lequeux, C. Ryter, M. G\'erin, 
Astron. Astrophys. 354 (2000) L87.

\hang\noindent
[61] C. Moutou, K. Sellgren, L. Verstraete, A. L\'eger, 
Astron. Astrophys. 347 (1999) 949.

\hang\noindent
[62] J.D. Bregman, L.J. Allamandola, A.G.G.M. Tielens, F.C. Witteborn, 
D.M. Rank, D. Wooden, 1986, in
Summer School on Interstellar Processes: Abstracts of Contributed Papers,
ed. D.J. Hollenbach, H.A. Thronson, 87.

\hang\noindent
[63] J. Bregman, 1989, in
Interstellar Dust, Internat. Astron. Union Symp. 135,
ed. L.J. Allamandola, A.G.G.M. Tielens
(Dordrecht: Kluwer Academic), 109.

\hang\noindent
[64] C. Joblin, P. Boissel, A. L\'eger, L. d'Hendecourt, D. D\'efourneau, 
Astron. Astrophys. 299 (1995) 835.

\hang\noindent
[65] Y.J. Pendleton, S.A. Sandford, 
L.J. Allamandola, A.G.G.M. Tielens, K. Sellgren, 
Astrophys. J. 437 (1994) 683.

\hang\noindent
[66] J.E. Chiar, A.G.G.M. Tielens, D.C.B. Whittet, 
W.A. Schutte, A.C.A. Boogert, 
D. Lutz, E.F. van Dishoeck, M.P. Bernstein, 
Astrophys. J. 537 (2000) 749.

\hang\noindent
[67] W.A. Schutte, K.A. van der Hucht, D.C.B. Whittet, A.C.A. Boogert, 
A.G.G.M. Tielens, P.W. Morris, J.M. Greenberg, P.M. Williams, 
E.F. van Dishoeck, J.E. Chiar, Th. de Graauw, 
Astron. Astrophys. 337 (1998) 261.

\hang\noindent
[68] D.A. Beintema, M.E. van den Ancker, F.J. Molster, L.F.B.M. Waters,
A.G.G.M. Tielens, C. Waelkens, T. de Jong, Th. de Graauw, K. Justtanont,
I. Yamamura, A. Heras, F. Lahuis, A. Salama,
Astron. Astrophys. 315 (1996) L369.

\hang\noindent
[69] L.J. Allamandola, S.A. Sandford, A.G.G.M. Tielens, T.M. Herbst, 
Astrophys. J. 399 (1992) 134.

\hang\noindent
[70] R.S. Lewis, T. Ming, J.F. Wacker, E. Anders, E. Steel
Nature 326 (1987) 160.

\hang\noindent
[71] E. Anders, E. Zinner, 
Meteoritics 28 (1993) 490.

\hang\noindent
[72] H.W. Kroto, J.R. Heath, S.C. O'Brien, R.F. Curl, R.E. Smalley, 
Nature 318 (1985) 162.

\hang\noindent
[73] H.W. Kroto, M. Jura, 
Astron. Astrophys. 263 (1992) 275.

\hang\noindent
[74] A. Goeres, E. Sedlmayr, 
Astron. Astrophys. 265 (1992) 216.

\hang\noindent
[75] R.P.A. Bettens, E. Herbst, 
Astrophys. J. 478 (1997) 585.

\hang\noindent
[76] A. Scott, W.W. Duley, G.P. Pinho, 
Astrophys. J. 489 (1997) L193.

\hang\noindent
[77] L. Becker, T.E. Bunch, 
Meteoritics 32 (1997) 479.

\hang\noindent
[78] D. Heymann, 
Astrophys. J. 489 (1997) L111.

\hang\noindent 
[79] T.P. Snow, C.G. Seab, 
Astron. Astrophys. 213 (1989) 291.

\hang\noindent 
[80] W.B. Somerville, J.G. Bellis, 
Mon. Not. Roy. Astron. Soc. 240 (1989) 41P.

\hang\noindent
[81] B.H. Foing, P. Ehrenfreund, 
Nature 369 (1994) 296.

\hang\noindent
[82] B.H. Foing, P. Ehrenfreund, 
Astron. Astrophys. 317 (1997) L59.

\hang\noindent
[83] G.H. Herbig, 
Ann. Rev. Astron. Astrophys. 33 (1995) 19.

\hang\noindent
[84] J. Fulara, M. Jakobi, J.P. Maier, 
Chem. Phys. Letters 211 (1993) 227.

\hang\noindent
[85] R.D. Oudmaijer, G. Busfield, J.E. Drew, 
Mon. Not. Roy. Astron. Soc. 291 (1997) 797.

\hang\noindent
[86] J.E. Chiar, A.J. Adamson, D.C.B. Whittet, 
Astrophys. J. 472 (1996) 665.

\clearpage
\centerline
{\bf Figure Captions}
\medskip

{\bf Figure 1}--- 
The cosmic abundances of elements from Cox [1], presented as
log$_{10}$ of the elemental abundance relative to hydrogen versus
the atomic number of the element.  Only elements with abundances
greater than 10$^{-6}$ relative to hydrogen are illustrated.
 
\bigskip
{\bf Figure 2}--- The ISOCAM + CVF spectrum of the reflection nebula
NGC 7023 from Cesarsky et al. [2], plotted as relative flux density
versus frequency in cm$^{-1}$.
The spectral resolution is $\nu / \Delta \nu$ = 40.
Five of the six interstellar emission features (IEFs) can be seen,
at 790, 890, 1160, 1300, and 1610 cm$^{-1}$
(6.2, 7.7, 8.6, 11.3, and 12.7 $\mu$m).
 
\bigskip
{\bf Figure 3}--- IRAS broad-band photometric observations 
({\it filled squares}) of
$R(12/{\rm total})$, the ratio of the 
flux in the IRAS 12 $\mu$m filter to the total infrared flux,
plotted against $T_{\rm eff}$(star), the 
effective stellar temperature of the star which
illuminates each reflection nebula, from Sellgren et al. [43].
Error bars are $\pm$1-$\sigma$.
Upper limits are 3-$\sigma$.
The curves are the predicted values of $R(12/{\rm total})$ from
Pizagno et al. (2000, in preparation), 
derived by convolving the UV and visible
absorption curves of neutral naphthalene ({\it dashed curve}),
from Salama \& Allamandola [45],
and ionized naphthalene ({\it solid curve}), from
F. Salama (1999, private communication),
with theoretical energy distributions of stars as a function of
$T_{\rm eff}$(star).

\bigskip
{\bf Figure 4}--- Scaled ISOCAM + CVF spectra of three reflection nebulae,
vdB 17 (NGC 1333; {\it dotted line}), vdB 59 (NGC 2068; 
{\it thick solid line}), and vdB 133 ({\it thin solid line}),
plotted versus wavelength in $\mu$m, from Uchida et al. [44, 46].
The spectral resolution is $\lambda / \Delta \lambda$ = 40.
The stars which illuminate each reflection
nebula respectively have 
$T_{\rm eff}$ = 11,000 K for vdB 17, 19,000 K for vdB 59, and
6,800 + 12,000 K for vdB 133 (which is illuminated by a binary star).

\bigskip
{\bf Figure 5}--- Ratio of the 5.5 -- 9.75 $\mu$m flux 
to the 10.25 -- 14.0 $\mu$m
flux versus the UV radiation field, $G_0$, from Uchida et al. [46].
All data are derived from ISOCAM + CVF spectra.
Different nebular positions are illustrated for
vdB 17 ({\it filled diamonds}; [46]),
and vdB 59 ({\it filled circles}; [46]).
Results at a single nebular position are shown for
vdB 133 ({\it open circle}; [46]),
NGC 7023 ({\it cross}; from Cesarsky et al. [2]),
and $\rho$ Oph ({\it open square}; from Boulanger et al. [15]).
Error bars are $\pm$1-$\sigma$.  

\bigskip
{\bf Figure 6}--- ISOCAM + CVF spectra of two reflection nebulae,
plotted versus wavelength in $\mu$m, from Uchida et al. [46].
The right panel is vdB 59 (NGC 2068), while the left panel is
vdB 17 (NGC 1333).  Spectra at different nebular positions are labeled
by their values of the UV intensity, $G_0$.
The spectral resolution is $\lambda / \Delta \lambda$ = 40.
Note the broadening of the 7.7 $\mu$m IEF at the lowest $G_0$ values
in vdB 17, which is unexplained by current models.

\bigskip
{\bf Figure 7}--- FWHM in $\mu$m
of the 6.2, 8.6, and 7.7 $\mu$m IEFs as a function
of the UV radiation field, $G_0$, within vdB 17, from Uchida et al. [46].
Data are derived from ISOCAM + CVF spectra.
Error bars are $\pm$1-$\sigma$.  No correction has been made
for the instrinsic spectral resolution of
the instrument ($\lambda / \Delta \lambda$ = 40).

\bigskip
{\bf Figure 8}--- Spectra at 7.0 -- 8.7 $\mu$m
of NGC 7023, plotted versus wavelength in $\mu$m, 
observed with ISO's Short Wavelength
Spectrometer (SWS) at $\lambda / \Delta \lambda$ = 1800 by Moutou et al.
[61].  Solid vertical lines mark the laboratory positions
of C$_{60}^+$ bands [84].
The data labeled ``(up -- down)/2'' provide an estimate of the noise
of the spectrum.
The emission line in NGC 7023 at 8.02 $\mu$m is the 0--0 S(4) line of
molecular hydrogen.
The SWS responsivity
calibration is shown at the bottom of the plot,
offset for clarity.

\bigskip
{\bf Figure 9}--- ({\it top}) Optical depth, plotted versus wavelength in
$\mu$m, of a possible aromatic 
absorption feature toward the protostar Mon R2/IRS--3, embedded in
a molecular cloud, from ground-based spectra
($\lambda / \Delta \lambda$ = 1000)
obtained by Sellgren et al. [19]
at the United Kingdom Infrared Telescope Facility. 
({\it middle}) Optical depth, plotted versus wavelength in
$\mu$m, of a possible aromatic 
absorption feature toward
the enigmatic Galactic Center source GCS3, which
is obscured primarily by the diffuse interstellar medium, from
Short Wavelength Spectrometer spectra 
($\lambda / \Delta \lambda$ = 1000)
from ISO observed by
Chiar et al. [66].  
({\it bottom}) Normalized intensity of the 3.3 $\mu$m IEF toward
the reflection nebula NGC 7023 (Sellgren et al. 2000, in preparation), from
the Short Wavelength Spectrometer on ISO ($\lambda / \Delta \lambda$ = 1000).
The emission line at 3.234 $\mu$m is the 1--0 O(5) 
line of molecular hydrogen.

\bigskip
{\bf Figure 10}--- The optical depth of the 3.47 $\mu$m absorption
band in molecular clouds,
possibly due to ``diamond''-like carbon, plotted versus the optical
depth of 9.7 $\mu$m silicate absorption, a refractory grain
component.  
Data are from Brooke et al. [22] ({\it filled circles}),
Sellgren et al. and Brooke et al. [18, 21] ({\it open circles}),
and Chiar et al. [82] ({\it crosses}).

\bigskip
{\bf Figure 11}--- The optical depth of the 3.47 $\mu$m absorption
band in molecular clouds, 
possibly due to ``diamond''-like carbon, plotted versus the optical
depth of 3.1 $\mu$m H$_2$O ice absorption, a volatile grain
component.
Data are from Brooke et al. [22] ({\it filled circles}),
Sellgren et al. and Brooke et al. [18, 21] ({\it open circles}),
and Chiar et al. [82] ({\it crosses}).

\clearpage
\pagestyle{empty}

\begin{figure}
\plotfiddle{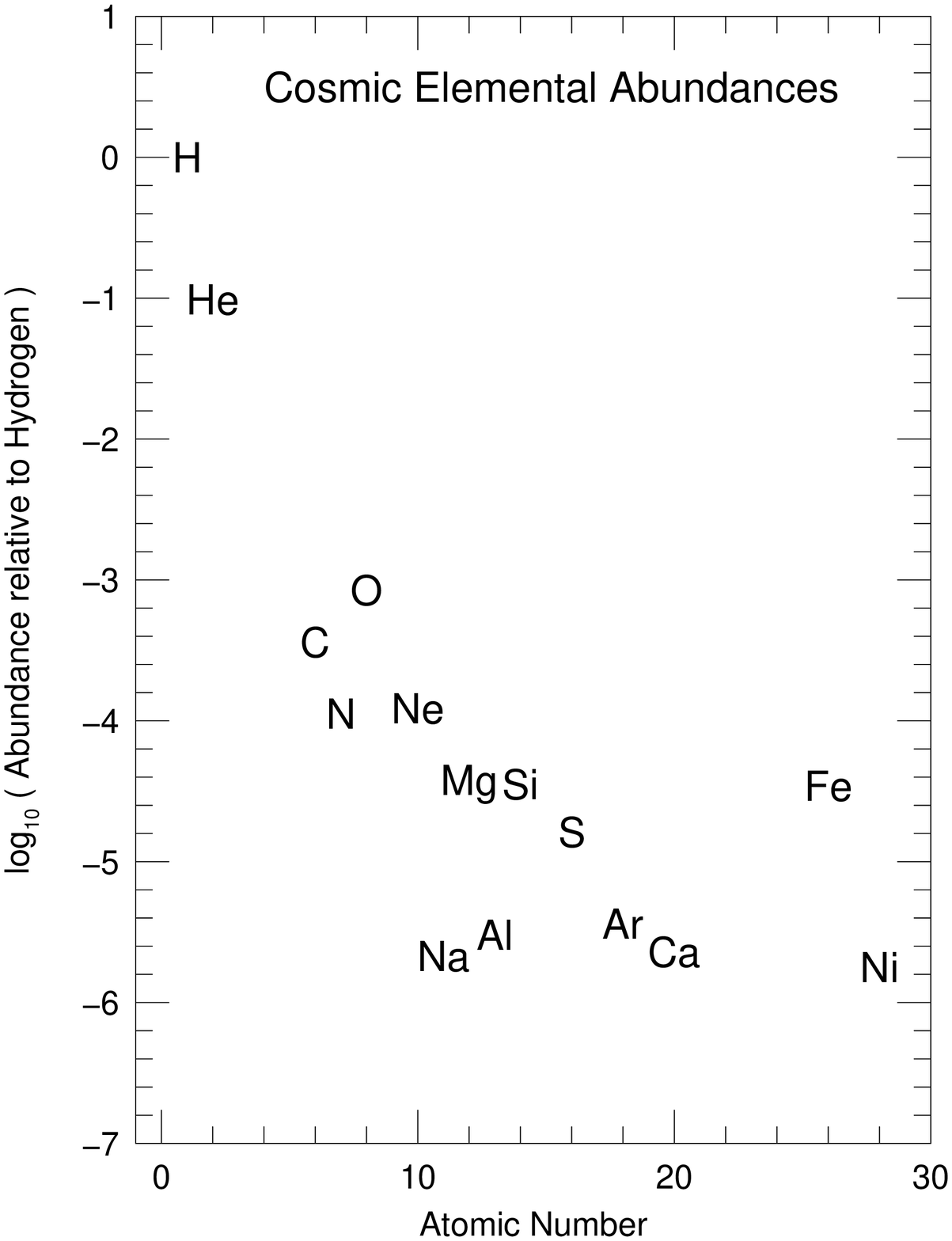}{9.0in}{0}{100}{100}{-324}{0}
\end{figure}

\begin{figure}
\plotfiddle{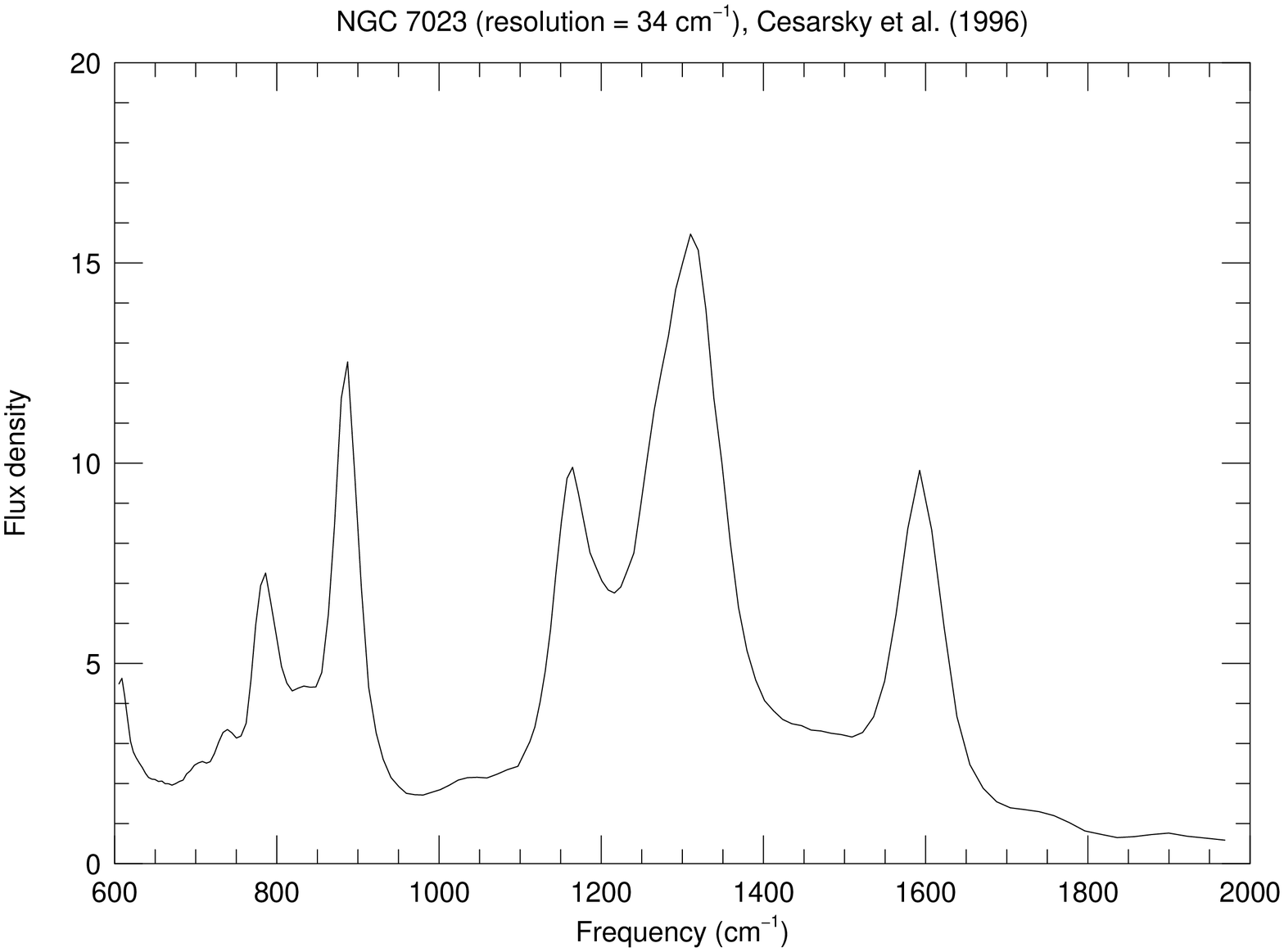}{9.0in}{90}{100}{100}{324}{0}
\end{figure}

\begin{figure}
\plotfiddle{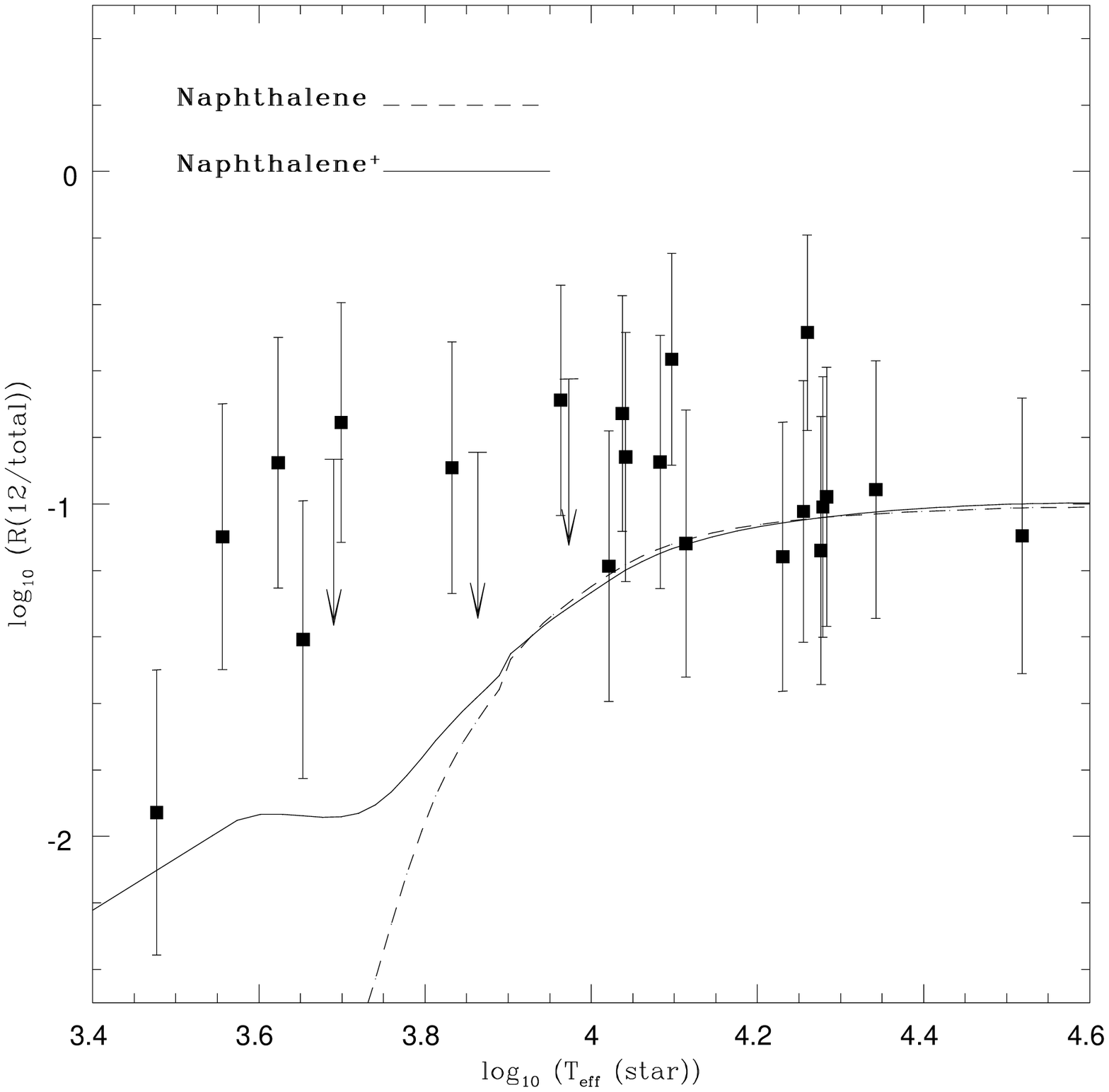}{9.0in}{0}{100}{100}{-324}{0}
\end{figure}

\begin{figure}
\plotfiddle{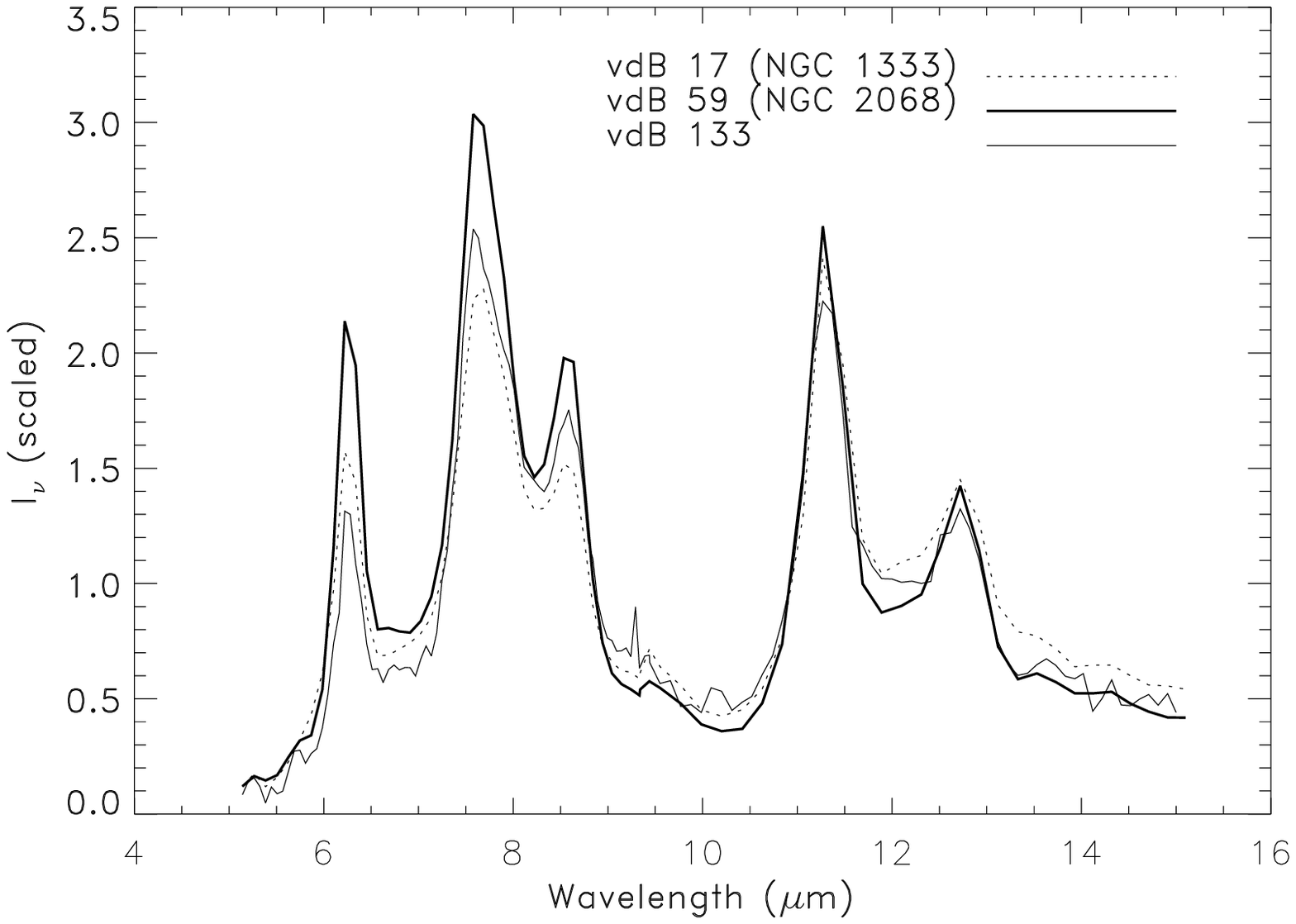}{9.0in}{90}{100}{100}{324}{0}
\end{figure}

\begin{figure}
\plotfiddle{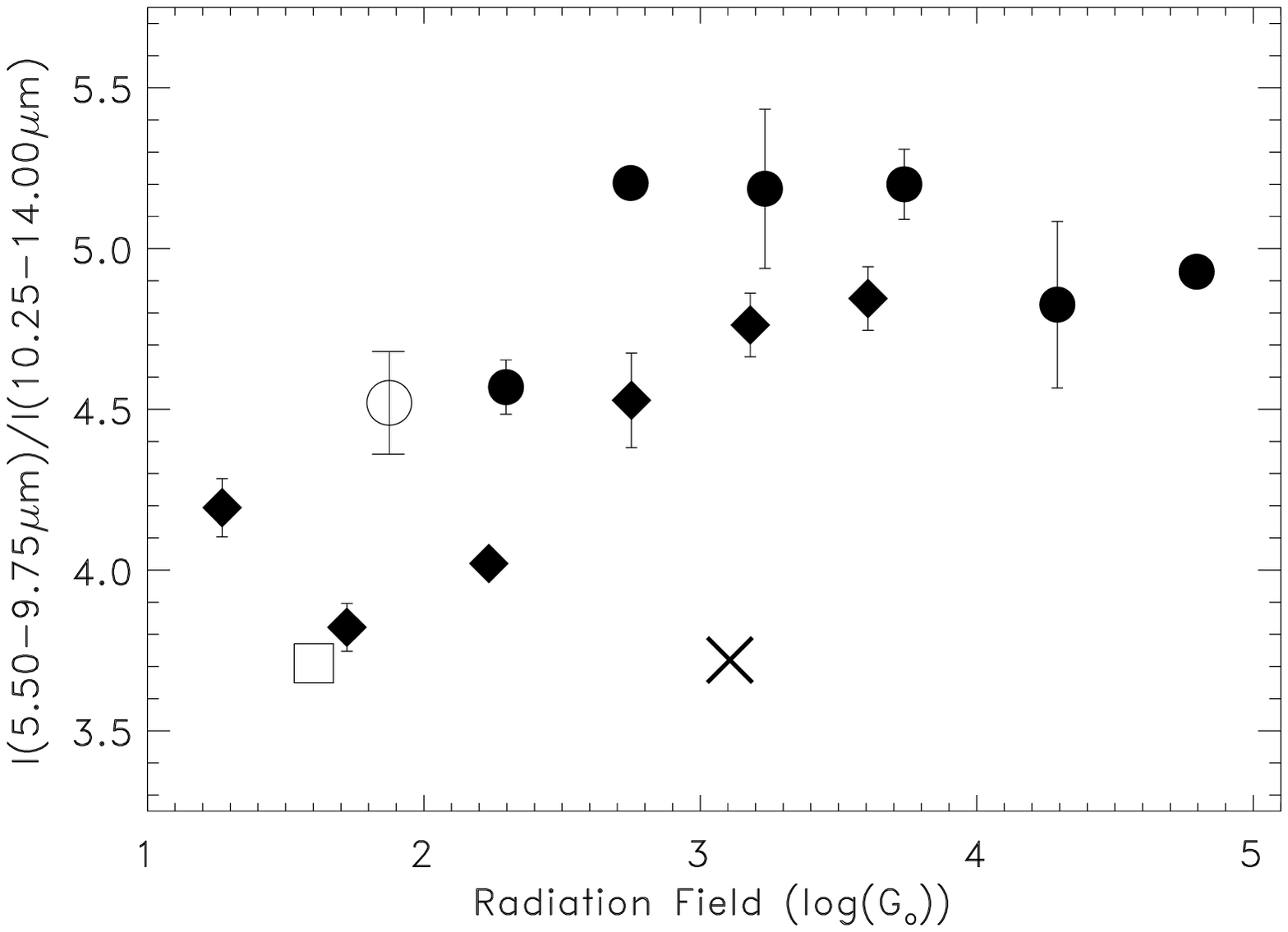}{9.0in}{90}{100}{100}{324}{0}
\end{figure}

\begin{figure}
\plotfiddle{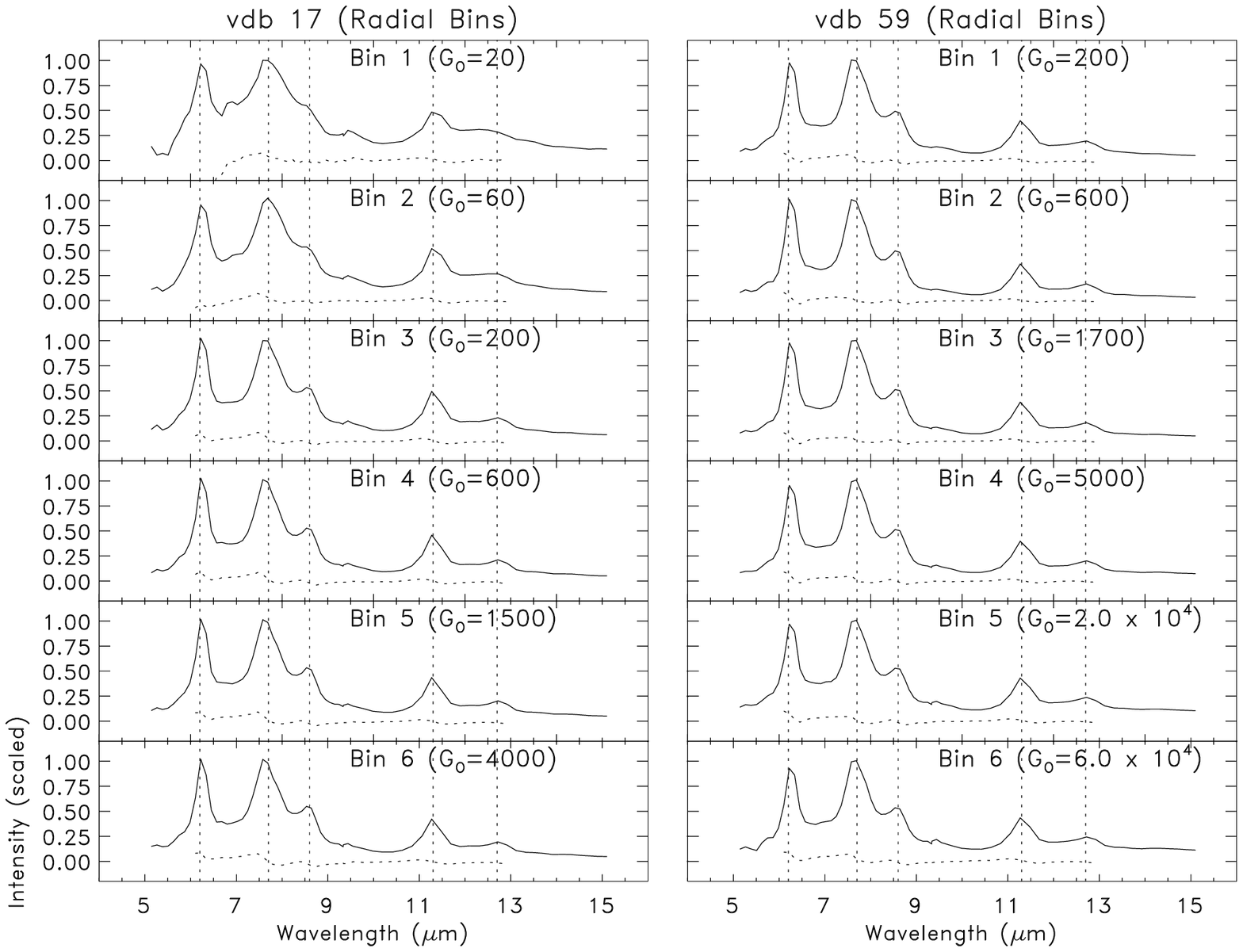}{9.0in}{0}{100}{100}{-324}{0}
\end{figure}

\begin{figure}
\plotfiddle{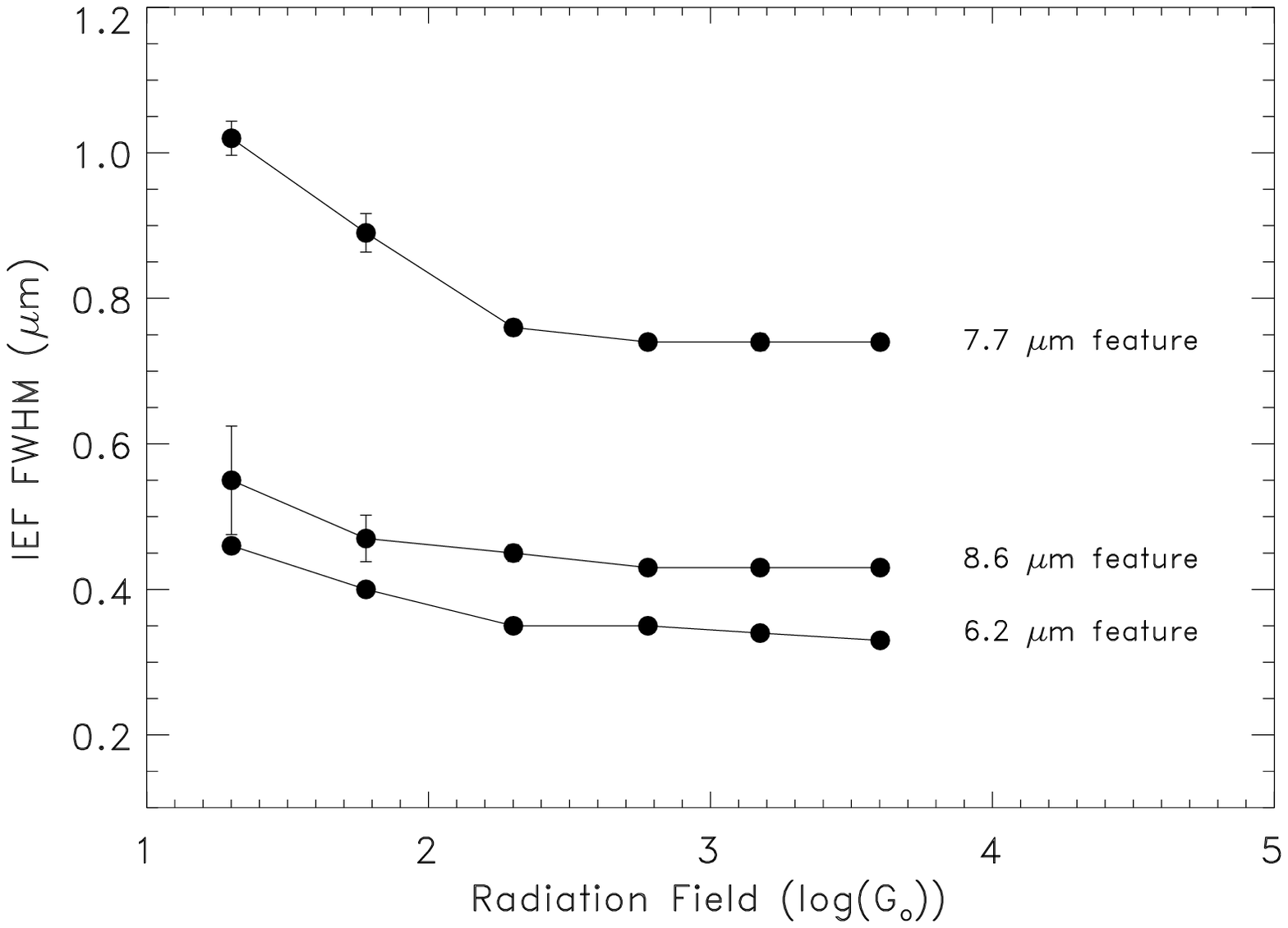}{9.0in}{90}{100}{100}{324}{0}
\end{figure}

\begin{figure}
\plotfiddle{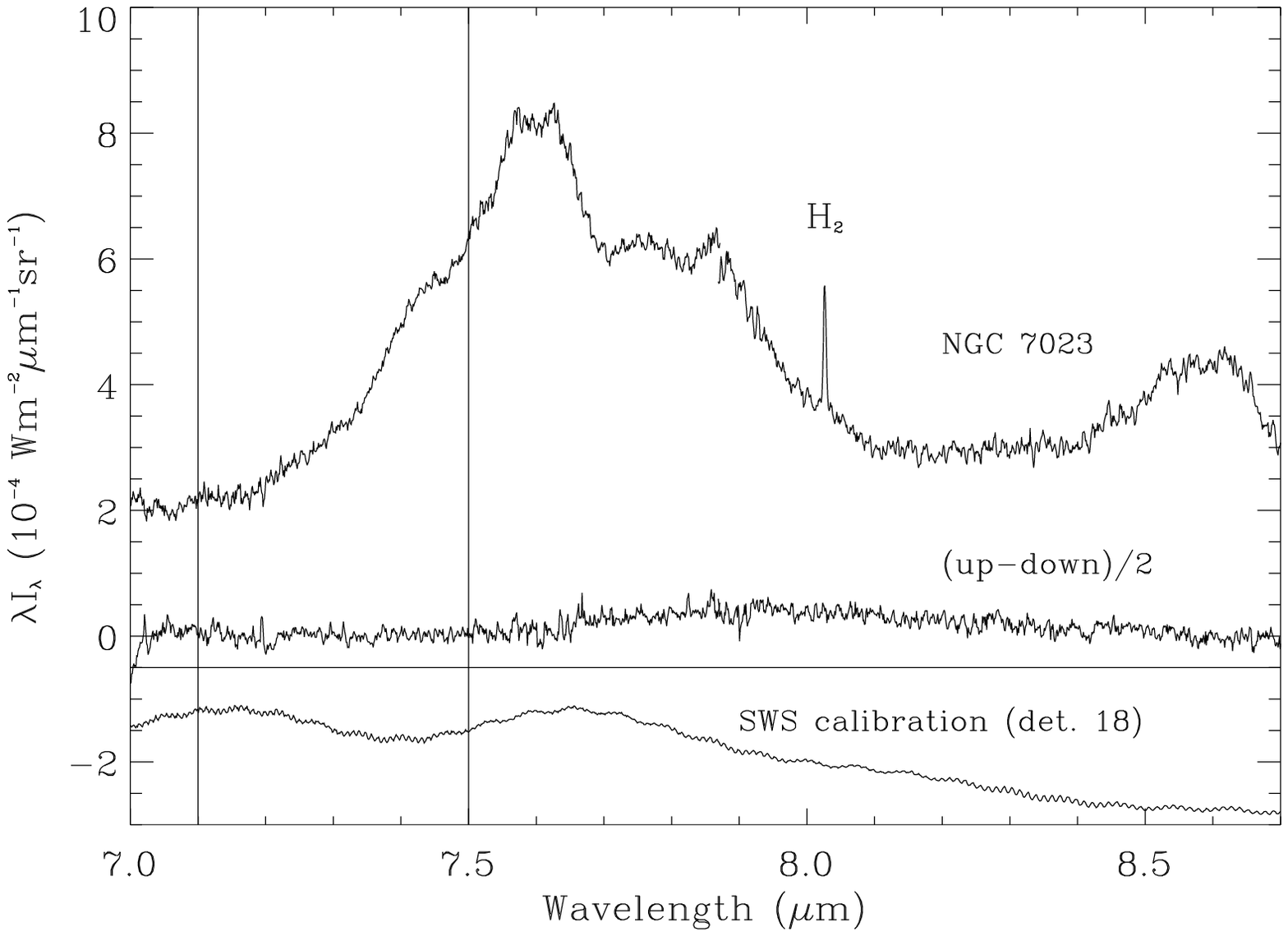}{9.0in}{0}{100}{100}{-324}{0}
\end{figure}

\begin{figure}
\plotfiddle{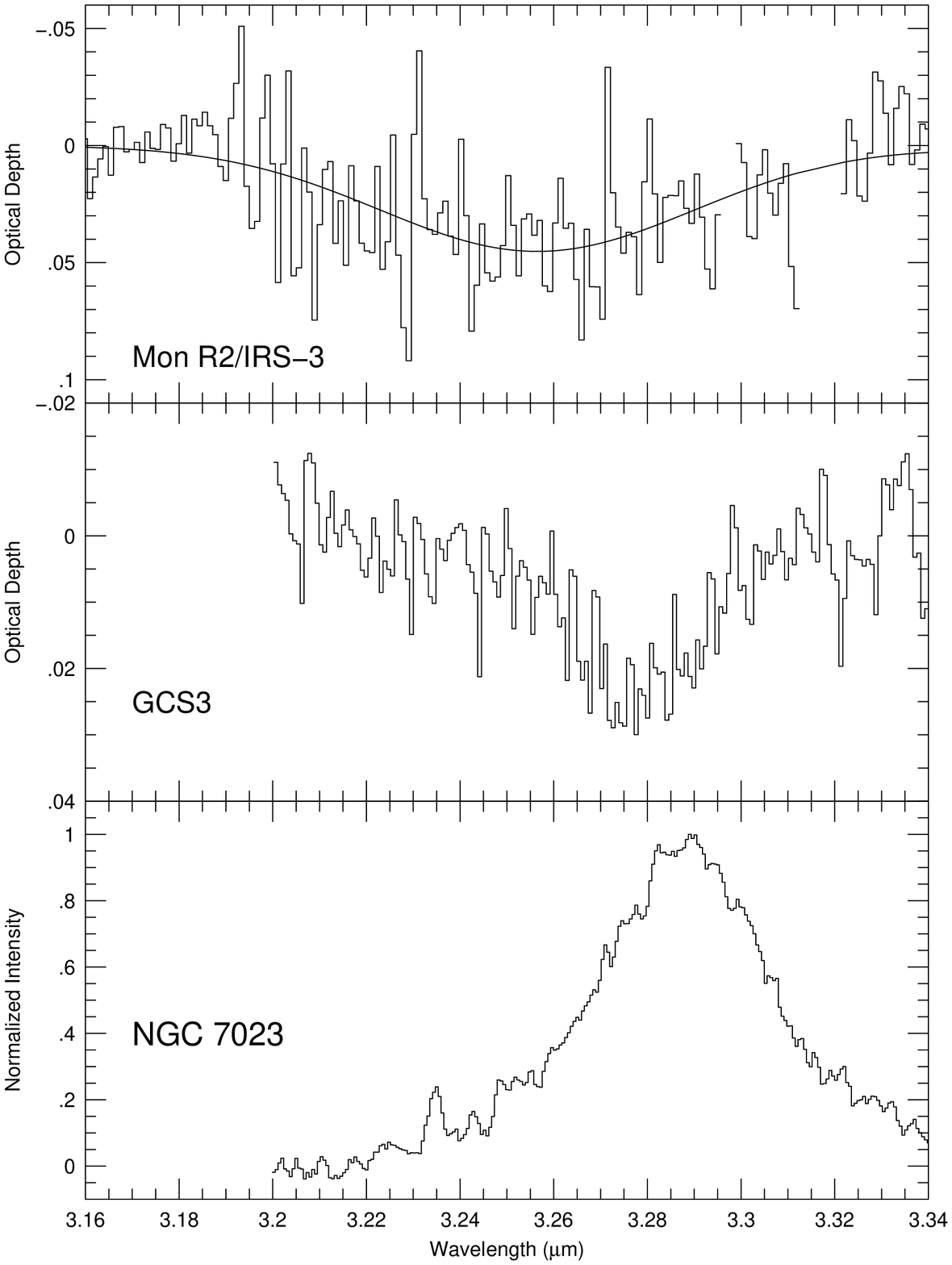}{9.0in}{0}{100}{100}{-324}{0}
\end{figure}

\begin{figure}
\plotfiddle{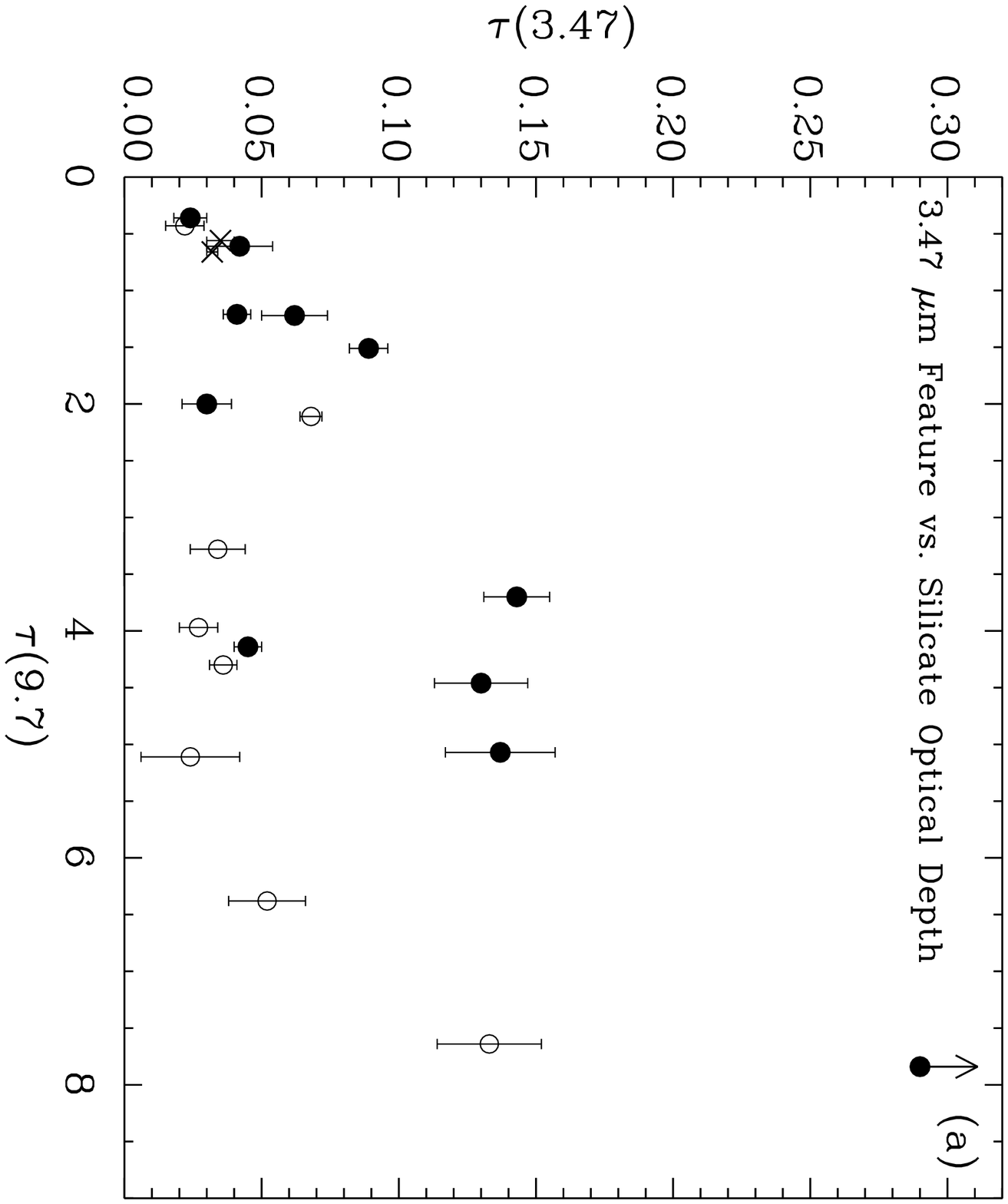}{9.0in}{0}{100}{100}{-324}{0}
\end{figure}

\begin{figure}
\plotfiddle{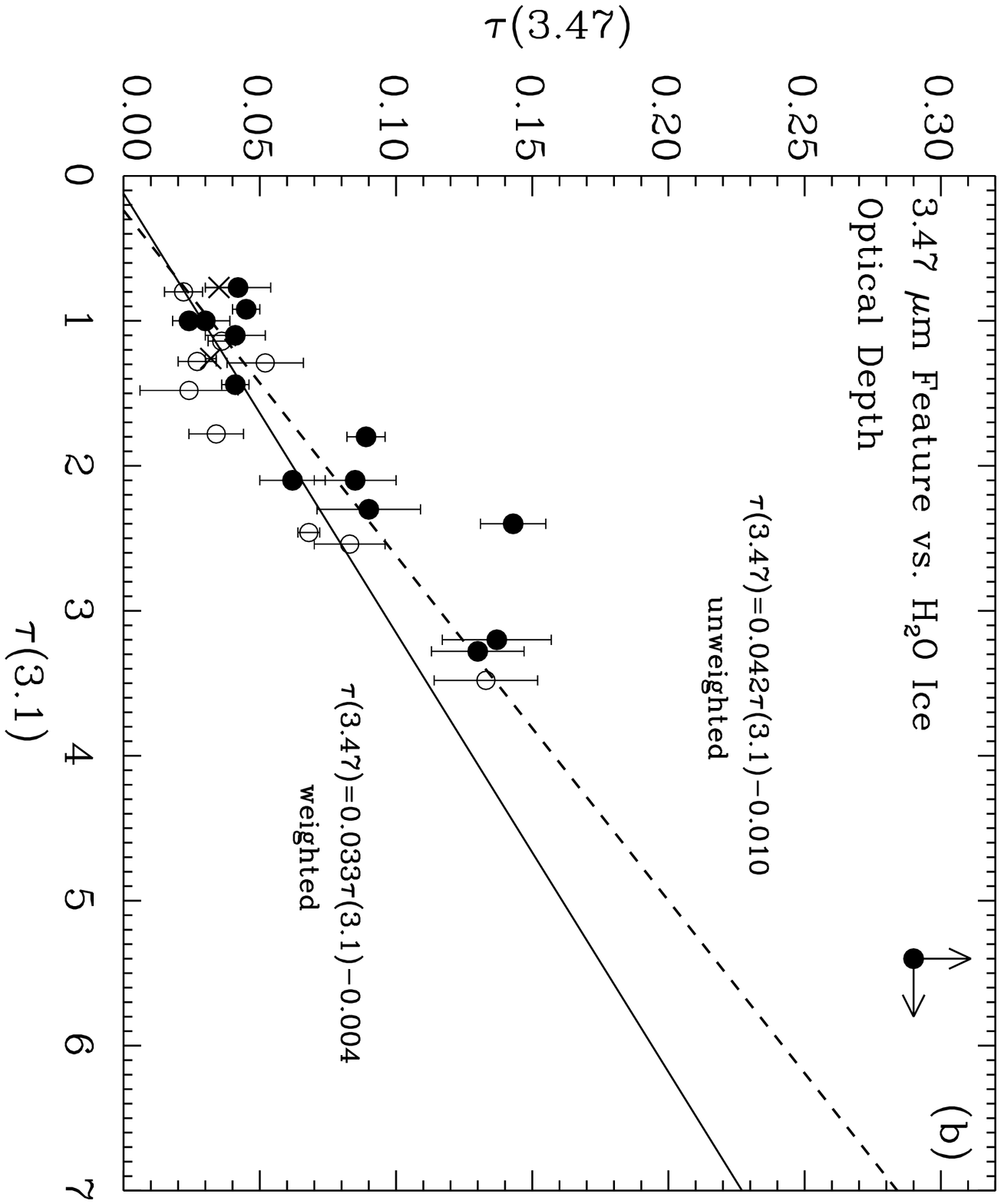}{9.0in}{0}{100}{100}{-324}{0}
\end{figure}

\end{document}